\documentclass[12pt]{article}
\newcommand{\tr}[1]{\,{\rm tr}\,#1\,}

\begin{document}
\title{ Consistent noncommutative quantum gauge theories?}
\author{A.A.Slavnov \\ Steklov Mathematical
Institute, \\Gubkina st. 8, GSP-1, 117966 Moscow, Russia}  \maketitle

\begin{abstract}

A new noncommutative model invariant with respect to $U(1)$ gauge group
 is proposed. The model is free of nonintegrable infrared
 singularities. Its commutative classical limit describes a free
 scalar field. Generalization to $U(N)$ models is also
 considered.

\end{abstract}

\section{Introduction}

Perturbative aspects of noncommutative quantum theories were recently
 a subject of numerous investigations.
These studies revealed some peculiar feathures of noncom-\\mutative
models.  Noncommutativity introduces naturally nonlocality of interaction,
which serves as an ultraviolet regulator. However the regularization is
not complete and ultraviolet divergencies do not disappear completely.
Planar diagrams of a noncommutative theory require renormalization similar
to the procedure used in commutative models. Nonplanar diagrams become
ultraviolet convergent due to the presence of phase factors, however
the corresponding integrals have pole singularities in external
momenta leading to infrared divergency of higher order diagrams
(\cite{Fi}, \cite{KW}, \cite{MRS}, \cite{ABK}, \cite{GKW}).  In
particular these infrared singularities are present in noncommutative
$U(1)$ gauge theory, and although planar diagrams may be renormalized in a
gauge invariant way \cite{MS}, the model is inconsistent
(\cite{Ha}, \cite{MST}, \cite{ABKR}).

Another unusual property of noncommutative gauge theories is related to
the fact that noncommutative $SU(N)$ algebra is not closed and one is
forced to consider $U(N)$ models which include a $U(1)$ sector and
hence are also inconsistent (\cite{AA}, \cite{BS}).

One could try to cure this desease by introducing nonlocal counterterms
which cancel the infrared singularities. However it would
lead to a drastic modification of the original action and nobody was able
to prove that such a procedure may be carried out in a consistent and
gauge invariant way.

Experience obtained in commutative theories suggests that appearance of
diver-\\gencies, which cannot be removed by renormalizing charges, masses
and wave func-\\tions is a signal that the underlying classical theory is
not complete and should be modified in such a way that possible divergent
structures have the same form as the terms present in the classical
action.

Motivated by this observation we propose a modified noncommutative $U(1)$
invariant action which does not lead to infrared divergencies. All
divergencies are ultraviolet and may be removed by a standard
renormalization procedure. The classical theory in the limit
when the noncommutativity parameter $(\xi)$ tends to zero
 reduces, contrary to naive
expectations, not to free electrodynamics, but to a free scalar field
theory. A similar procedure may be applied to $U(N)$ noncommutative gauge
models, where the commutative classical limit describes $SU(N)$
vector bosons and $U(1)$ scalar particle.

\section{Noncommutative gauge invariant models.}

We start by reminding the basic facts about the conventional
noncommutative $U(1)$ theory.

The model is described by the action
\begin{equation}
S= \int d^4x \{- \frac{1}{4}F_{\mu \nu}F_{\mu \nu} \}
\label{1}
\end{equation}
\begin{equation}
F_{\mu \nu}= \partial_{\mu}A_{\nu}- \partial_{\nu}A_{\mu}+ig[A_{\mu}*
 A_{\nu}]
\label{2}
\end{equation}
The star product is defined as follows
\begin{equation}
f(x)* g(x)= \exp \{i \xi \theta_{\mu \nu} \partial_{\mu}^x
 \partial_{\nu}^y \}f(x)g(y)_{y=x}
\label{3}
\end{equation}
where $\theta_{\mu \nu}$ is a real antisymmetric matrix and $\xi$ is a
 noncommutativity parameter. In the limit $ \xi \rightarrow 0$ the action
 (\ref{1}) obviously reduces to the free electromagnetic action.

 The gauge transformations look similar to nonabelian gauge
 transformations
 \begin{equation}
\delta A_{\mu}= \partial_{\mu} \epsilon-ig(A_{\mu}* \epsilon-
\epsilon* A_{\mu})
 \label{4}
 \end{equation}
Note that although for a general skewsymmetric matrix $\theta_{\mu \nu}$
the interaction (\ref{1}) is nonlocal, models with $\theta_{i0}=0$
 introduce only spatial nonlocality and the standard Hamiltonian formalism
 may be applied. In what folows we assume that $\theta_{i0}=0$ and
 Hamiltonian formalism may be used. Without loss of generality one
 may take $ \theta_{12}=- \theta_{21}=1; \theta_{13}=
 \theta_{23}=0$.

One sees that the $U(1)$ noncommutative theory is nonabelian, and the
 Feynman rules look similar to the usual Yang-Mills theory. In particular
 Faddeev-Popov ghosts, parametrizing $ \det(\partial_{\mu}D_{\mu})$, where
 $D_{\mu}$ is the covariant derivative, are present. The free propagators
 coincide with the propagators of Yang-Mills theory, and the vertex
 functions are obtained substituting Lie algebra structure
 constants by the phase factors. For example the three point gauge vertex
 with momenta $p, q, k$ and indices $\mu, \nu, \rho$ looks as follows
 \begin{equation}
 2ig \sin(\xi p \tilde{q})[(p-q)_{\rho} \delta_{\mu
 \nu}+(q-k)_{\mu} \delta_{\nu \rho}+(k-p)_{\nu} \delta_{\mu \rho}]
 \label{5}
 \end{equation}
 Here we use the notation $\tilde{p}_{\mu}= \theta_{\mu \nu}p_{\nu}$.

 The gauge field polarization operator has ultraviolet divergent part
 corresponding to the planar diagrams, and the convergent nonplanar
 part, which contains the term singular at $p=0$. Explicit calculation
 gives \begin{equation} \Pi_{\mu \nu}(p)= \frac{g^2}{2 \pi^2}
 \frac{\tilde{p}_{\mu}\tilde{p}_{\nu}}{\xi^2(\tilde{p}^2)^2}+ \ldots
 \label{6} \end{equation}
 where $\ldots$ denotes less singular terms. One sees that $\Pi_{\mu \nu}$
 has a pole singularity at $p=0$ and the limit
 $\xi \rightarrow 0$ does not exist. The diagrams which have several
 insertions of $\Pi_{\mu \nu}$ into gauge field lines are infrared
 divergent.

 Similar singularities appear in the three point function, which looks as
 follows
 \begin{equation}
 \Gamma_{\mu \nu \rho}(p,q) \sim \cos(\xi p \tilde{q})
 \{ \frac{ \tilde{p}_{\mu} \tilde{p}_{\nu}  \tilde{p}_{\rho}}
{\xi(\tilde{p}^2)^2}+ sym \}+ \ldots
 \label{7}
 \end{equation}
 where $\ldots$ again stands for less singular terms and $sym$ means
  symmetrization
 \begin{equation}
 p \rightarrow q,\mu \rightarrow \nu;\quad p \rightarrow -(p+q), \mu
  \rightarrow \rho; \quad q \rightarrow -(p+q), \nu \rightarrow \rho.
 \label{8}
 \end{equation}

In general infrared pole singularities arise in the diagrams which in the
  absence of phase factors would be quadratically or linearly ultraviolet
 divergent. Logarihmically divergent diagrams produce only logarithmic
 infrared singularities which do not spoil integrability. In the
 commutative case gauge invariance prevents the appearance of linear and
 quadratic divergencies, but in the noncommutative theory they do appear.
 The only possible exception known so far is presented by supersymmetric
 gauge theories (\cite{SJ}, \cite{Z}, \cite{J-J}).

To avoid infrared divergencies one may try to subtract nonlocal
counterterms (\ref{6}, \ref{7}). However the meaning of such subtraction
 is not clear, as it does not correspond to renormalization of any
 parameter present in the original Lagrangian and the subtraction
 procedure is ambigous. Moreover, as was mentioned above, nobody proved
 that such subtraction can be done in a consistent and gauge invariant
 way.

 We consider the appearance of singular terms proportional to $
   \tilde{p}_{\mu}A_{\mu}$ as a signal that original action must be
 modified to include the terms of this type.

The action (\ref{1}) is not
 the only gauge invariant expression one can write in the noncommutative
 $U(1)$ theory. The most general gauge invariant action, which
 corresponds to a power counting renormalizable theory, posesses passive
 Lorentz invariance (i.e. is invariant if the tensor $\theta_{\mu \nu}$
 also undergoes Lorentz transformations), and introduces nonlocality only
 via star product has a form \begin{equation} S= \int d^4x \{-
 \frac{1}{4}F_{\mu \nu}F_{\mu \nu}+ \beta \lambda(x) \theta_{\mu
 \nu}F_{\mu \nu}(x)+ \gamma(\theta_{\mu \nu}F_{\mu \nu}(x))^2 \} \label{9}
 \end{equation}

 Here $\beta$ and $\gamma$ are arbitrary parameters and the
 Lagrange multiplier $\lambda(x)$ trans-\\forms according to adjoint
 representation of the gauge group.

 We choose  $ \beta=1$. Then obviously the last term is irrelevant and one
 may put $ \gamma=0$.  The action (\ref{9}) describes a
 constrained system and to study its physical content one has to formulate
 the Hamiltonian dynamics.  A natural requirement for a noncommutative
 theory is the condition that in the limit $\xi \rightarrow 0$ the Lorentz
 invariance is restored. The commutative limit of the action (\ref{9}) is 
  \begin{equation} 
 S_0= \int d^4x \{- \frac{1}{4}(\partial_{\mu}A_{\nu}- 
 \partial_{\nu}A_{\mu})^2+ \lambda(x) \tilde{\partial}_iA_i(x)
 \label{10} 
\end{equation}
We remind that we consider the case when the only nonzero 
elements of the matrix $\theta_{\mu \nu}$ are $\theta_{12}=- 
\theta_{21}=1$. In this case $\sum_i \tilde{\partial}_i^2= 
\sum_{i=1,2}\partial_i^2= \tilde{\partial}^2$.

At first sight the Lorentz invariance is broken even in the commutative 
limit. The proper Hamiltonian analysis shows however that the 
action(\ref{10}) describes a usual free scalar field.

Let us rewrite the equation (\ref{10}) as the action of a generalized
Hamiltonian system:
\begin{equation} 
S_0= \int d^4x \{p_i \dot A_i- \frac{p_i^2}{2}- 
\frac{1}{4}(\partial_iA_j-\partial_jA_i)^2+A_0 \partial_ip_i+ \lambda 
\tilde{\partial}_iA_i \}
\label{11} 
 \end{equation} 
 To fix the gauge we choose the Coulomb condition $\partial_iA_i=0$. Apart 
 from the first class constraint $\partial_ip_i=0$ the 
action (\ref{10}) includes the additional constraint 
$\tilde{\partial}_iA_i=0$. The commutator of this constraint with the 
Hamiltonian is different from zero:
\begin{equation} 
[ \int dx \frac{p_i^2}{2}, \theta_{ij} \partial_i A_j]= \tilde{\partial}_i 
p_i 
\label{12} 
\end{equation} 
That means the secondary constraint must be included and the complete 
action looks as follows
\begin{eqnarray}
S_0= \int d^4x \{p_i \dot A_i- \frac{p_i^2}{2}- \frac{1}{4}(\partial_iA_j- 
\partial_jA_i)^2+\nonumber\\
A_0 \partial_i p_i+ \lambda(x) \tilde{\partial}_iA_i+ \mu(x) 
\tilde{\partial}_ip_i 
\label{12a}
\end{eqnarray}

Let us parametrize the field $A_i$ and momentum $p_i$ as follows
\begin{eqnarray}
A_i= \partial_i \chi+ \tilde{\partial}_i \psi+ 
\epsilon^{ijk} \tilde{\partial}^{-2} \partial_j \tilde{\partial}_k \phi 
 \nonumber\\ p_i= \partial_i p_{\chi}+ \tilde{\partial}_i p_{\psi}+ 
 \epsilon^{ijk} \tilde{\partial}^{-2} \partial_j \tilde{\partial}_k 
p_{\phi} \label{13} \end{eqnarray} The Coulomb gauge condition and the 
constraint $\tilde{\partial}_iA_i=0$ insure that $\chi= \psi=0$.  The 
remaining constraints nullify the corresponding momenta.

After solution of the constraints and gauge condition the eq.(\ref{12a}) 
may be written in the following form 
\begin{equation}
S_0= \int d^4x \{p_{\phi} \dot \phi- \frac{p_{\phi}^2}{2}- 
 \frac{\partial_i\phi \partial_i \phi}{2}\}
\label{14} 
\end{equation}
which is the Lorentz invariant action for the scalar field. Contrary to 
 naive expecta-\\tions the limit $\xi \rightarrow 0$ of the modified 
electromagnetic action describes a scalar field.

Now we turn to the analysis of infrared singularities. We ignore 
 possible local determinants which appear due to the second class 
 constraints, assuming that some gauge invariant regularization (e.g. 
 dimensional) is used, in which these factors are absent. The Feynman 
 rules differ from the usual noncommutative 
 $U(1)$ theory by the presence of propagators of $\lambda$-fields, mixed 
 propagators $ \lambda A_{\mu}$ and the new vertex $g 
 \lambda[A_{\mu}*A_{\nu}] \theta_{\mu \nu}$. The corresponding elements 
 of diagram technique are:

The propagator $\lambda, \lambda$
 \begin{equation} 
k^2 \tilde{k}^{-2}
\label{l5} 
\end{equation} 

The propagator $\lambda, A_{\mu}$
\begin{equation} 
\tilde{k}_{\mu}(\tilde{k})^{-2}
\label{16} 
\end{equation} 

The vertex $\lambda, A^2$
\begin{equation} 
ig \sin(\xi p \tilde{q}) \theta_{\mu \nu}
 \label{18} 
\end{equation} 

The propagator of the Yang-Mills field is also modified. In the diagonal 
Feynman gauge it is
 \begin{equation}
\frac{1}{k^2}(g^{\mu \nu}- \frac{\tilde{k}_{\mu} 
\tilde{k}_{\nu}}{\tilde{k}^2})
 \label{19}
 \end{equation}

 We start with the one loop polarization operator. To study the leading 
 singularities we may put the external momentum equal to zero everywhere 
except for the phase factors. Thus we have
\begin{equation}   
\Pi_{\mu \nu}^{sing}(p)= \int d^4k \sin^2(\xi p \tilde{k})P_{\mu 
\nu}(k)
 \label{20} \end{equation} 
where $P_{\mu \nu}$ is a rational function with the dimention $k^{-2}$. To 
separate the infrared singular contribution one presents the phase 
factor as 
\begin{equation} 
\sin^2(\xi p \tilde{k})= \frac{1}{2}(1- \cos(2\xi p \tilde{k}))
\label{21} 
\end{equation} 
The constant term corresponds to the planar contribution and produces 
ultraviolet divergency which is removed by the usual wave function 
renormalization, whereas the term proportional to $ \cos(2\xi p 
\tilde{k})$ gives the infrared singular contribution \begin{equation} 
\Pi_{\mu \nu}^{sing}(p) \sim \frac{Ag^{\mu 
\nu}\tilde{p}^2+B\tilde{p}^{\mu} \tilde{p}^{\nu}}{ \xi^2(\tilde{p}^2)^2} 
 \label{22} 
\end{equation} 
The polarization operator must satisfy ST-identities, which in this case 
reduce to transversality condition
\begin{equation}
p_{\mu}p_{\nu} \Pi_{\mu \nu}^{sing}=0 \rightarrow A=0
\label{23} 
\end{equation} 
Hence the pole singularity is proportional to $ \tilde{p}_{\mu} 
\tilde{p}_{\nu}$.  Recalling that the free propagator (\ref{19}) of the 
Yang-Mills field $A_{\mu}$ is transversal with respect to $ 
\tilde{p}_{\mu}$, we conclude that the infrared singularity is 
irrelevant if the polarization operator is connected to other part of 
a diagram by the gauge field propagator.
 The next possible singular term is $ \sim \ln(p^2)$ and 
does not lead to nonintegrable singularity.

In our model there is also a mixed propagator $A_{\mu} \lambda$. 
 Due to this mixing the singular part of $\Pi_{\mu \nu}$ may contribute to 
 the amplitude with four external gauge field lines, obtained by 
connecting the polarization operator $\Pi_{\mu \nu}$ with the 
vertices (\ref{18}) by the mixed propagators. The corresponding 
contribution is proportional to 
\begin{equation}
\frac{\sin^2( \xi p \tilde{q})}{(\tilde{p}^2)^2 \xi^2}
\label{23a}
\end{equation}
and does not lead to infrared divergencies at small $p$. Note that the 
limit $\xi \rightarrow 0$ is also nonsingular.

There are also one-loop polarization operators $\Pi_{\mu}(p)$, 
corresponding to diagrams with one external $A_{\mu}$-line and one 
$\lambda$-line, and  $\Pi(p)$, corresponding to diagrams with two 
$\lambda$-lines. All these diagrams in the absence of phase factors 
diverge at most logarithmically and therefore are infrared safe.

 However if the the polarization operator of $\lambda$-field, $\Pi(p)$ 
required renormalization, that would mean that the action (\ref{9}) was not 
complete and new counterterms $\sim \lambda^2$ have to be introduced.

Moreover the complete $\lambda$-field propagator may include subsequent 
 insertions of several polarization operators $\Pi_{\mu \nu}, \Pi_{\mu}, 
\Pi$ connected by the mixed propagator $A_{\mu} \lambda$. For example
\begin{equation}
D_{\lambda \lambda}(p)=D_{\lambda A_{\mu}}(p) \Pi_{\mu \nu}(p)D_{A_{\nu} 
\lambda}(p) \Pi(p)D_{\lambda A_{\rho}}(p) \Pi_{\rho 
\sigma}(p)D_{A_{\sigma} \lambda}(p) \label{23b} \end{equation} Obviously 
a diagram containing such propagator may produce infrared singularities 
 due to accumulation of the $\Pi_{\mu \nu}$ poles.

Let us study the infrared behaviour of $\Pi(p)$ more closely. In the 
 lowest order the polarization operator $\Pi(p)$ is equal to
\begin{eqnarray}
\Pi(p)= \int d^4k[\sin(p \tilde{k} \xi) \theta_{\mu \nu}(g_{\mu \alpha}- 
 \frac{\tilde{k}_{\mu} \tilde{k}_{\alpha}}{\tilde{k}^2})k^{-2} 
\times\nonumber\\(g_{\nu \beta}- \frac{(\tilde{p}+ 
\tilde{k})_{\nu}(\tilde{p}+ 
\tilde{k})_{\beta}}{(\tilde{p}+\tilde{k})^2})(p+k)^{-2} \sin(p \tilde{k} 
\xi) \theta_{\alpha \beta}] \label{23c} \end{eqnarray} Performing the 
multiplication explicitely we can rewrite this expression in the form 
\begin{eqnarray}
\Pi(p)= \int d^4k \sin^2(p \tilde{k} \xi)[\theta_{\alpha \beta}- 
\frac{k_{\beta} \tilde{k}_{\alpha}}{\tilde{k}^2}+ 
\frac{(p+k)_{\alpha}(\tilde{k}+ \tilde{p})_{\beta}}{(\tilde{k}+ 
\tilde{p})^2}+\nonumber\\ 
\frac{(\tilde{p}k) \tilde{k}_{\alpha}(\tilde{k}+\tilde{p})_{\beta}}
{\tilde{k}^2(\tilde{p}+\tilde{k})^2}]k^{-2}(p+k)^{-2} \theta_{\alpha 
\beta}
\label{23d}
\end{eqnarray}
According to our choice of $\theta_{\mu \nu}$ in this equation $\alpha, 
 \beta=1,2$. The most singular terms vanish after summation over $\alpha, 
\beta$, providing absolute convergence of the integral (\ref{23d}).
\begin{equation}
\Pi(p)= \int d^4k \sin^2(p \tilde{k} \xi) 
\frac{(\tilde{p}k)^2}{\tilde{k}^2(\tilde{p}+\tilde{k})^2}
k^{-2}(p+k)^{-2}
\label{23e}
\end{equation}
To estimate the infrared behaviour of $\Pi(p)$ let us take 
 $|p_1|=|p_2|=p$. Rescaling the integration variables 
$p \xi k \rightarrow x$ we get
\begin{equation}
\Pi(p)=p^4 \xi^2f(p^2 \xi)
\label{23f}
\end{equation}
where the function $f(p^2 \xi)$ has a logarithmic singularity at the 
origin $f(p^2 \xi)_{p \sim 0} \sim \ln(p^2 \xi)$. Therefore up to 
 logarithmic corrections $\Pi(p)$ vanishes at $p=0$ as $p^4$. It 
 compensates the infrared singularity of the operator $\Pi_{\mu \nu}$ and 
guarantees the infrared convergence. 

A general polarization operator $\Pi(p)$ may be analyzed in a similar way.
It may be presented in the following form
\begin{eqnarray}
\Pi(p)= \int d^4k \ldots d^4s \{\sin(p \tilde{k} \xi) \theta_{\alpha 
 \beta}(g_{\alpha \mu}- \frac{\tilde{k}_{\mu} 
\tilde{k}_{\alpha}}{\tilde{k}^2})(g_{\nu \beta}- \frac{(\tilde{p}+ 
\tilde{k})_{\nu}(\tilde{p}+ \tilde{k})_{\beta}}{(\tilde{p}+ \tilde{k})^2}) 
\times\nonumber\\
\Pi_{\mu \nu \rho \sigma}(p,k,s)(g_{\rho \lambda}- 
 \frac{\tilde{s}_{\rho} \tilde{s}_{\lambda}}{\tilde{s}^2})(g_{\sigma 
 \kappa}- \frac{(\tilde{s}+ \tilde{p})_{\sigma}(\tilde{s}+ 
\tilde{p})_{\kappa}}{(\tilde{s}+ \tilde{p})^2}) \theta_{\lambda \kappa} 
\sin(p \tilde{s} \xi) \} \label{23g} \end{eqnarray} We consider the case 
when the outer vertices are connected with the internal part of the 
 diagram by the gauge field propagators.  There are also the diagrams 
 where some of these propagators are replaced by the mixed propagators 
 $A_{\mu} \lambda$. They are considered in a similar way and we shall not 
 present the analysis here. The function $\Pi_{\mu \nu \rho \sigma}$ in 
 the eq.(\ref{23g}) may be separated into parts symmetric and 
 antisymmetric with respect to $\mu \nu$ and $\rho \sigma$. The 
 antisymmetric parts are proportional to $\theta_{\mu \nu}$ and 
 $\theta_{\rho \sigma}$ respectively. 
 Performing the summation over all indices one sees that the first 
 nonvanishing term is proportional to $p^2$. 
 Therefore the integral is absolutely convergent and to study its 
behaviour at $\xi \sim 0$ we may put $\xi \sim 0$ in the integrand.
In this way one sees that the $\Pi(p, \xi)$ and its first derivative over 
$\xi$ vanish at $\xi=0$. Possible asymptotics of $\Pi(p, \xi)$ at $\xi 
\sim 0$ have the form 
\begin{equation} 
\Pi(p, \xi)_{\xi \sim 0}=\xi^n \ln^m(\xi)
\label{23i}
\end{equation}
 Therefore $\Pi(p, \xi)$ for small $\xi$ is proportional to $\xi^2 
\ln^m(\xi)$.  By dimensinal reasons at $p \rightarrow 0$, $\Pi(p) \sim p^4 
\xi^2 \ln^m(p^2 \xi)$, in accordance with the lowest order result.
 
 Insertion of the mixed polarization operator $\Pi_{\mu}(p)$ does not 
 change our analysis. This operator vanishes at $p=0$ as $|p|$. At the 
 same time the mixed propagator $A_{\mu} \lambda$ has a singularity $\sim 
 |p|^{-1}$. So the product is not singular. It is important to note that 
 to have two insertions of $\Pi_{\mu \nu}(p)$ which might produce 
 nonintegrable infrared singularity one needs at least one insertion of 
 the operator $\Pi(p)$, which cancels the singularity.

Now we turn to the study of three point gauge field vertex. It satisfies 
the ST-identity, which we take in the original form (\cite{Sl}):
\begin{eqnarray} 
<A_{\mu}(x)A_{\nu}(y) \partial_{\rho}A_{\rho}(z)>=
<\partial_{\mu}M^{-1}_{xz}A_{\nu}^b(y)>+\nonumber\\
g<A_{\mu}(x)M^{-1}_{xz}A_{\nu}(y)>+( \mu \rightarrow \nu, x \rightarrow 
 y).
\label{24} 
\end{eqnarray} 
Here $M^{-1}_{xy}$ is the Green function of ghost field in the external 
gauge field.
 
 We start again with the one-loop diagrams. To pass to the proper vertex 
  function we must amputate external propagators, which include both 
  $A_{\mu} A_{\nu}$-propagators and mixed propagators $A_{\mu} \lambda$. 
  However mixed propagators and mixed proper vertex functions do not have 
  pole singularities and being interested in the leading singular terms, 
 we may drop them. The gauge-ghost vertices and ghost propagators which 
 enter the r.h.s. of eq.(\ref{24}) also have no pole singularities as in 
 the absence of phase factors the corresponding integrals diverge 
 logarithmically.  Keeping only the terms which may have pole 
 singularities we rewrite the eq(\ref{24}) in the form \begin{equation} 
\frac{(p+q)_{\rho}}{(p+q)^2}(g_{\mu \alpha}- \frac{\tilde{p}_{\mu} 
 \tilde{p}_{\alpha}}{\tilde{p}^2})(g_{\nu \beta}- \frac{\tilde{q}_{\nu} 
 \tilde{q}_{\beta}}{\tilde{q}^2}) \Gamma^{1, 
 sing}_{\alpha \beta \rho}(p,q)=0
\label{24a} 
 \end{equation} 
where $\Gamma^{1, sing}_{\mu \nu \rho}$ is the pole singular part of the 
one-loop proper vertex function. In deriving this equation we used the 
transversality of the free gauge field propagators with respect to 
$\tilde{p}_{\mu}$ and established earlier fact that the singular part of 
$ \Pi_{\mu \nu}(p)$ is proportional to 
$\tilde{p}_{\mu} \tilde{p}_{\nu}$. By the same reasonings as above the 
singular part of $\Gamma^1_{\mu \nu \rho}(p,q)$ depends only on $ 
\tilde{p}, \tilde{q}$.  The only possible structure which has a proper 
symmetry and dimension, and satisfies the identity (\ref{24a}) is 
\begin{equation} \Gamma^{1, sing}_{\mu \nu \rho}(p,q) \sim \{ 
\frac{\tilde{p}_{\mu} \tilde{p}_{\nu} 
 \tilde{p}_{\rho}}{\xi^2|\tilde{p}|^4}+ (p \rightarrow q)+(p \rightarrow 
 -(p+q)) \} \label{25} 
 \end{equation}
 Due to transversality of the free gauge field propagator $\Gamma^{1, 
sing}_{\mu \nu \rho}$ does not contribute to the vertex function with 
  three external gauge lines. It might give a nonzero contribution to the 
  diagram with four external gauge lines obtained by connecting 
  $\Gamma^{1, sing}_{\mu \nu \rho}$ with the vertex (\ref{18}) by the 
  mixed propagators $A_{\mu} \lambda$. However as in the case of the 
  two point polarization operator this diagram does not produce 
  infrared singularities. 
  Proper vertex functions with at least one external $\lambda$-line in 
  the absence of the phase factors diverge logarithmically and do 
  not produce infrared divergencies.
  
To analyze higher loop diagrams one should perform carefully the
renormalization and check if our estimates remain valid. It is not done 
in the present paper. Assuming that renormalization does not introduce new 
problems we may basically repeat our arguments for arbitrary multiloop 
diagrams.

 The singular part of the polarization operator again may be 
calculated by taking external momentum equal to zero everywhere except for 
the phase factors.  (Of course we assume that necessary ultraviolet 
subtractions of divergent subgraphs are done in accordance with 
$R$-operation). Therefore the singular part of polarization operator at 
arbitrary order depends only on $\tilde{p}$.  Gauge invariance and 
dimensional reasons fix the form of the singular part to $\tilde{p}_{\mu} 
\tilde{p}_{\nu}|\tilde{p}|^{-4}$.  Hence the arguments given above to 
prove the absence of infrared singularities may be applied directly.

The singular part of the three point function depends only on $\tilde{p}, 
 \tilde{q}$ and by gauge invariance must satisfy the identity (\ref{24}). 
An analogue of the eq.(\ref{24a}) for a singular part of $\Gamma^n 
_{\mu \nu \rho}$ will now include the terms
\begin{equation} 
\frac{(p+q)_{\rho}}{(p+q)^2} \sum_{m,l=1}^{n-1}D_{\mu \alpha}^m(p)D_{\nu 
\beta}^l(q) \Gamma_{\mu \nu \rho}^{n-m-l, sing}(p,q) 
 \label{26} 
 \end{equation} 
  We proved that the two point Green functions and proper 
one-loop three-point Green functions have no pole singularities at zero 
momenta.  Assuming that it is true for 
all $m<n$, we see that these terms do not produce singular contributions 
to $\Gamma^n_{\mu \nu \rho}(p,q)$ and its structure is also given by the 
eq.(\ref{25}). It completes the induction.

A similar modification allows to formulate a consistent noncommutative 
 $U(N)$ gauge theory. The standard noncommutative $U(N)$ Yang-Mills action 
is
\begin{equation}
S=\int d^4x \tr[- \frac{1}{8}F_{\mu \nu}*F_{\mu \nu}]
  \label{28} 
 \end{equation} 
 where
 \begin{equation} 
F_{\mu \nu}= \partial_{\mu}A_{\nu}- 
 \partial_{\nu}A_{\mu}+g[A_{\mu}*A_{\nu}]
 \label{29} 
 \end{equation} 
 and $A_{\mu}$ belongs to $U(N)$ Lie algebra. Due to noncommutativity of 
the star product this action mixes the $U(1)$ and the $SU(N)$ gauge 
bosons, leading to UV/IR mixing analogous to the pure $U(1)$ case. The one 
loop diagrams in this theory were analyzed in (\cite{AA}). It 
appears that the planar diagrams in this theory may be 
renormalized in a gauge invariant way. Nonplanar 
diagrams with the $U(1)$ boson external lines are infrared 
singular, whereas the nonplanar diagrams with only $SU(N)$ 
boson external lines do not exibit infrared singularities.

The infrared pole singularities may be eliminated in analogy with the 
$U(1)$ case.
 
 Let us consider the modified $U(N)$ action:
 \begin{equation} 
 S= \int d^4x \tr[- \frac{1}{8}F_{\mu \nu}*F_{\mu \nu}]+ \lambda(x) 
\tr[\theta_{\mu \nu}F_{\mu \nu}] \label{30} 
\end{equation} 
 where the Lagrange multiplier $\lambda(x)$ belongs to the adjoint 
 representation of the $U(1)$ group, and $F_{\mu \nu}$ is the 
$U(N)$ curvature tensor.
 
 The free action consists of the usual $SU(N)$ part and modified $U(1)$ 
 action considered above. So the spectrum includes vector $SU(N)$ bosons 
and the scalar particle assosiated with the $U(1)$ group. The analysis of 
infrared singularities given above was based on the gauge invariance, 
power counting and the explicit form of $U(1)$ propagators. Therefore 
it may be applied directly to the diagrams with at least one 
external $U(1)$ line and leads to the same conclusion. These diagrams are 
free of infrared pole singularities. The one loop diagrams with only 
$SU(N)$ external lines were shown to be infrared safe. If this 
property holds at higher loops, the noncommutative $U(N)$ theory has the 
same infrared properties as the commutative one and is renormalizable. 
 
\section{Discussion}

In this paper I wanted to show that consistent 
 noncommutative quantum gauge theories free of infrared singularities may 
 exist even in the absense of supersymmetry.  The crucial observation 
 which allows to construct such models is a possibility to describe by 
  noncommutative gauge models not only vector but also scalar fields. 
 
Several questions may be raised in this connection. 
 
 We did not 
 consider carefully the ultraviolet renormalization of the theory. 
 It seems very plausible that the ultraviolet 
 renormalization preserves the invariance of the model and does not 
 change our estimates of asymptotics, but it would be good to demonstrate 
 it explicitely.
 
 We concentrated in this paper on pole singularities, which lead to 
 infrared divergency. However the logarithmic singularities, which do not 
 cause infrared problems, may be present. Commutative limit of 
 quantum theory deserves further investigation.
 
 Our proof of existence of the noncommutative quantum $U(N)$ model 
  assumed the absence of infrared singularities in the diagrams with 
  pure $SU(N)$ external lines. To my knowleadge explicit proof of this 
  fact has been given for one-loop diagrams (\cite{AA}). Although it is 
  likely to be true for a general diagram, a careful study 
  would be useful.

 {\bf Acknowledgements.} \\
This work was initiated while the author was visiting the University of 
 Heraklyon. I thank E.Kiritsis and the theoreical group for hospitality. I 
 am grateful to D.Maison and J.Wess for hospitality at 
 Max-Planck-Institute in Munich and Humboldt Foun-\\dation for a generous 
 support. My thanks to A.Koshelev for helpful discussions.The work was 
 supported in part by Russian Basic Research Foundation under grant 
 02-01-00126, and grant for the support of leading scientific schools.  

\begin{thebibliography}{99}{\small \bibitem{Fi} T.Filk, Phys.Lett. B376 
 (1996) 53. \bibitem{KW} T.Krajewsky, R.Wulkenhaar, 
 Int.J.Mod.Phys. A15 (2000) 1011. 
  \bibitem{MRS} S.Minwalla, M.Van Raamsdonk, N.Seiberg, 
 JHEP 9906 (1999) 007. 
  \bibitem{ABK} I.Ya.Aref'eva, 
 D.M.Belov, A.S.Koshelev, Phys.Lett. B476 (2000) 431.  
\bibitem{GKW} H.Grosse, T.Krajewski, R.Wulkenhaar, hep-th/0001182
\bibitem{MS} C.P.Martin, D.Sanchez-Ruiz, Phys.Rev.Lett. 83 (1999) 476. 
 \bibitem{Ha} M.Hayakawa, Phys.Lett. B478 (2000) 394. 
  \bibitem{MST} A.Matusis, L.Susskind, N.Tombas, JHEP 0012 
 (2000) 002.  
\bibitem{ABKR} I.Ya.Aref'eva, D.M.Belov,  A.S.Koshelev, O.A.Rytchkov,  
Nucl.Phys. B406 (1993) 90.  \bibitem{AA} A.Armoni, Nucl.Phys. B593 (2001) 
 229.  \bibitem{BS} L.Bonora, M.Salizzoni, Phys.Lett. B504 (2001) 80.  
 \bibitem{SJ} M.M.Sheikh-Jabbari, JHEP 9905 (1999) 015.  \bibitem{Z} 
 D.Zanon, Phys.Lett.  B502 (2001) 265.  \bibitem{J-J} I.Jack, D.R.T.Jones, 
New J.Phys.31 (2001) 19.  \bibitem{Sl} A.A.Slavnov, 
Theor.Math.Phys.  10 (1972) 99.} \end{thebibliography}
  \end{document}